\documentstyle[osa,manuscript]{revtex}
\begin{document}
%
\title{Measurement of $B(D_s^+\rightarrow\mu^+\nu_\mu)
/B(D_s^+\rightarrow\phi\mu^+\nu_\mu)$
and Determination of the Decay Constant $f_{D_s}$}
%
\author{
Fermilab E653 Collaboration\\
K.~Kodama$^{(1)}$,
S.~Torikai$^{(1)}$,
N.~Ushida$^{(1)}$,
A.~Mokhtarani$^{(2),(a)}$,
V.S.~Paolone$^{(2)}$,
J.T.~Volk$^{(2),(a)}$,
J.O.~Wilcox$^{(2),(b)}$,
P.M.~Yager$^{(2)}$,
R.M.~Edelstein$^{(3)}$,
A.P.~Freyberger$^{(3),(c)}$,
D.B.~Gibaut$^{(3),(d)}$,
R.J.~Lipton$^{(3),(a)}$,
W.R.~Nichols$^{(3),(e)}$,
D.M.~Potter$^{(3)}$,
J.S.~Russ$^{(3)}$,
C.~Zhang$^{(3)}$,
Y.~Zhang$^{(3),(f)}$,
H.I.~Jang$^{(4)}$,
J.Y.~Kim$^{(4)}$,
B.R.~Baller$^{(5)}$,
R.J.~Stefanski$^{(5)}$,
K.~Nakazawa$^{(6)}$,
S.H.~Chung$^{(7)}$,
M.S.~Park$^{(7)}$,
I.G.~Park$^{(7)}$,
J.S.~Song$^{(7)}$,
C.S.~Yoon$^{(7)}$,
M.~Aryal$^{(8)}$,
N.W.~Reay$^{(8)}$,
R.A.~Sidwell$^{(8)}$,
N.R.~Stanton$^{(8)}$,
S.~Yoshida$^{(8)}$,
M.~Chikawa$^{(9)}$,
T.~Hara$^{(10)}$,
S.~Aoki$^{(11),(g)}$,
K.~Hoshino$^{(11)}$,
M.~Kobayashi$^{(11)}$,
M.~Komatsu$^{(11)}$,
M.~Miyanishi$^{(11)}$,
M.~Nakamura$^{(11)}$,
S.~Nakanishi$^{(11)}$,
K.~Niwa$^{(11)}$,
M.~Nomura$^{(11)}$,
K.~Okada$^{(11)}$,
H.~Tajima$^{(11),(h)}$,
K.~Teraoka$^{(11)}$,
J.M.~Dunlea$^{(12),(i)}$,
S.G.~Frederiksen$^{(12),(h)}$,
S.~Kuramata$^{(12),(j)}$,
B.G.~Lundberg$^{(12),(a)}$,
G.A.~Oleynik$^{(12),(a)}$,
K.~Reibel$^{(12)}$,
G.R.~Kalbfleisch$^{(13)}$,
P.~Skubic$^{(13)}$,
J.M.~Snow$^{(13)}$,
S.E.~Willis$^{(13),(k)}$,
K.~Nakamura$^{(14)}$,
T.~Okusawa$^{(14)}$,
T.~Yoshida$^{(14)}$,
H.~Okabe$^{(15)}$,
J.~Yokota$^{(15)}$,
N.~Ihara$^{(16)}$,
M.~Kazuno$^{(16)}$,
T.~Koya$^{(16)}$,
E.~Niu$^{(16),(l)}$,
S.~Ogawa$^{(16)}$,
H.~Shibuya$^{(16)}$,
S.~Watanabe$^{(16),(m)}$,
N.~Yasuda$^{(16)}$,
K.~Ehara$^{(17)}$,
K.~Horie$^{(17)}$,
Y.~Sato$^{(17)}$,
K.~Suzuki$^{(17)}$,
and
I.~Tezuka$^{(17)}$
}
\maketitle
%
\begin{abstract}
We have observed $23.2\pm6.0_{-0.9}^{+1.0}$ purely-leptonic 
decays of $D_s^+ \rightarrow \mu^+ \nu_\mu$
from a sample of muonic one prong decay events
detected in the emulsion target of Fermilab experiment E653.
Using the $D_s^+ \rightarrow \phi \mu^+ \nu_\mu$ yield measured 
previously in this experiment,
we obtain $B(D_s^+\rightarrow\mu^+\nu_\mu)
/B(D_s^+\rightarrow\phi\mu^+\nu_\mu)=0.16 \pm 0.06 \pm 0.03$.
In addition, we extract the decay constant
$f_{D_s}=194 \pm 35 \pm 20 \pm 14 \ MeV$.
\end{abstract}

%

1) Aichi University of Education, Kariya 448, JAPAN

2) University of California (Davis), Davis, CA 95616, USA

3) Carnegie-Mellon University, Pittsburgh, PA 15213, USA

4) Chonnam National University, Kwangju 500-757, KOREA

5) Fermi National Accelerator Laboratory, Batavia, IL 60510, USA

6) Gifu University, Gifu 501-11, JAPAN

7) Gyeongsang National University, Jinju 660-300, KOREA

8) Kansas State University, Manhattan, KS 66506, USA

9) Kinki University, Higashi-Osaka 577, JAPAN

10) Kobe University, Kobe 657, JAPAN

11) Nagoya University, Nagoya 464, JAPAN

12) The Ohio State University, Columbus, OH 43210, USA

13) University of Oklahoma, Norman, OK 73019, USA

14) Osaka City University, Osaka 558, JAPAN

15) Science Education Institute of Osaka Prefecture, Osaka 558, JAPAN

16) Toho University, Funabashi 274, JAPAN

17) Utsunomiya University, Utsunomiya 321, JAPAN

$^a$ Fermilab, Batavia, IL 60510, USA.

$^b$ Northeastern University, Boston, MA 02115, USA.

$^c$ CEBAF, 12000 Jefferson Avenue Newport News, VA 23606, USA.

$^d$ Virginia Polytechnic Institute and State University,
Blacksburg, VA 24060, USA.

$^e$ Westinghouse Electric Corp., Pittsburgh, PA 15230, USA.

$^f$ Pennsylvania State University, University Park,
PA 16802, USA.

$^g$ Kobe University, Kobe 657, JAPAN.

$^h$ University of Tokyo 7-3-1 Hongo, Bunkyou-ku, Tokyo 113 JAPAN.

$^i$ University of Rochester, Rochester, NY 14627, USA.

$^j$ Hirosaki University, Hirosaki 036, JAPAN.

$^k$ Northern Illinois University, DeKalb, IL 60115, USA.

$^l$ CERN, CH-1211, Geneva, SWITZERLAND.

$^m$ University of Wisconsin, Madison, WI 53706, USA.

\newpage

%

Leptonic decay of charged pseudoscalar mesons is described by the
annihilation of constituent quark and antiquark into a virtual 
$W$ boson. The rate for this process is proportional to the square 
of the pseudoscalar decay constant $f$, which measures the overlap
of quark and antiquark at zero separation. The constant $f$
gives absolute normalizations of numerous heavy-flavor 
transitions, including mixing, semi-leptonic and non-leptonic decays.
Hence it is highly desirable to determine the pseudoscalar decay 
constant more precisely in charm and $B$ physics.
A measurement of a purely-leptonic decay branching ratio,
unlike semi-leptonic or non-leptonic decays, is the most 
reliable way to extract the pseudoscalar decay constant, 
because it does not involve any QCD corrections.

There are three purely-leptonic decay modes for the $D_s^+$
(charge conjugate modes are implied throughout this paper):
electronic, muonic and tauonic. Among them, the muonic decay 
is the most accessible mode to determine the decay constant $f_{D_s}$.
The electronic decay is helicity suppressed by five orders of 
magnitude with respect to the muonic one. While the tauonic decay 
branching ratio is expected to be an order of magnitude larger than 
the muonic decay, the Q value for $D_s^+ \rightarrow \tau^+ \nu_\tau$ 
is small and hence reconstruction of this mode is extremely difficult 
because of the small decay angle in the lab frame.

The first measurement of the pseudoscalar decay constant of the charmed
strange meson, $f_{D_s}$, was carried out by the WA75 group in 
1993~[\cite{wa75}], using a transverse momentum spectrum of muons 
from $D_s^+$ leptonic decays observed in an emulsion target.
They obtained a value of $f_{D_s}=(232\pm45\pm20\pm48)\ MeV$.
CLEO~[\cite{cleo}] and BES~[\cite{bes}] also reported recent 
measurements of $f_{D_s}$, and ARGUS~[\cite{argus}] obtained $f_{D_s}$ 
by model dependent calculation in 1992. All of these results, however, 
suffer from large statistical and systematic errors, and hence the 
present measured value of $f_{D_s}$ is still not determined well 
enough to discriminate between different theoretical models.
In this paper, we report an independent measurement of $f_{D_s}$
from an analysis of the data from Fermilab experiment E653.

%

 The Fermilab emulsion-hybrid experiment E653 was designed to study
production and decay of heavy flavor particles by the direct
observation of decay vertices in the emulsion.
Data for this analysis were taken in a $600 GeV/c$ $\pi^-$ beam
during the second run of Fermilab E653. The trigger required both
a beam particle to interact in the target and a muon to penetrate
$3,900\ g/cm^2$ of absorber. This muon trigger provided an enriched
sample of semi-muonic decays of heavy flavor particles.  A total of 
$8.2\times10^6$ events, corresponding to $2.5\times10^8$ interactions, 
were recorded during this run.  Since the details of this experiment 
were described in another paper~[\cite{e653nim}], only a brief 
description of each detector element is given here.

The detector consisted of the emulsion target, tracking detector,
spectrometer, and muon detection system. Because of its unsurpassed 
spatial resolution, nuclear emulsion was chosen as an active target, 
and it was used both as a primary interaction target and as a decay 
vertex detector. The electronic detectors downstream of the target 
select events for the emulsion scanning and predict the interaction 
vertices and each track position in the emulsion.
 Tracks from interactions and decays in the emulsion were measured 
in the vertex silicon strip detectors (VSSDs).  The VSSDs consisted 
of 18 planes arranged in 6 groups of three readout coordinates, 
and the center part of the detector achieved a resolution transverse 
to the beam of $8.8 \mu m$. This detector also served as the upstream 
arm of the spectrometer. The spectrometer gave a momentum resolution of
$\sigma_p / p = \sqrt{ (0.01)^2 + (0.00023p)^2 }\ \ (p\ in\ GeV/c)$.
The most down stream element of the E653 detector array was the muon
identification system, which remeasured the direction and momenta
of muon candidates after the absorber with drift chambers and
an iron toroid.

%

To enrich the relative yield of events with heavy flavor particle 
decays and to reduce the emulsion scanning load, the transverse 
momentum of the triggered muon with respect to the beam axis, 
$P_{T\mu\ Beam}$, was required to be greater than $0.8 GeV/c$.
In total, 94,000 events passed this criterion from a initial sample 
of $8.2\times10^6$ events. This reduced sample contained a significant
number of background events from feedthrough muons. Pions or kaons 
could be misidentified as muons because they decayed muonically in 
flight with small angle deflections or from punch through in the iron 
absorber.  These events were rejected using the method described below.

In the emulsion analysis, the primary interaction vertices were 
visually located in the emulsion by a semi-automatic scanning 
microscope system using position predictions from the VSSDs. 
The efficiency for finding primary vertices was determined to be 
greater than 99\%. If the angle of an emulsion track from the primary 
vertex matched to within 1 mrad of the reconstructed muon track, 
the muon was identified as a feedthrough muon and the event was 
rejected. For the remaining events, a more detailed emulsion scanning 
technique was employed to find the origin of the muon.

Different decay search methods were applied for charged and neutral
decay vertices. To find charged decays, all emulsion tracks observed 
at the primary vertex which did not match to any track reconstructed 
by the VSSDs were followed towards the most downstream emulsion plates. 
To find neutral decays, the tracks reconstructed by the VSSDs
with no corresponding emulsion track observed at the primary vertex,
were located at the most downstream plate of the emulsion target.
Once found, these tracks were followed upstream through the emulsion
stack to their origin within the emulsion target to detect neutral 
decays. The detailed event location and vertex search methods are
described elsewhere~[\cite{scannim}].

 As a result of the above analysis, 1,193 charmed particle candidates 
were identified: 565 KINK (charged one prong) decays, 536 VEE (neutral 
two prong) decays, and 92 three or four prong decays.
Further cuts on the decay length ($>500 \mu m$) and the transverse 
momentum of the muon with respect to the parent particle direction 
($P_{T\mu}$ $>0.28 GeV/c$) were applied to the KINK and VEE decays 
to improve position resolution measurement and to reject strange 
particle decays.  Only VEEs with their vertices found in the emulsion 
were retained to avoid the large $P_{T\mu}$ error due to the 
intrinsically larger position errors of the VSSDs. The offline 
selection criteria and cuts are summarized in Table \ref{selection}.

%
%
\begin{center}
TABLE \ref{selection}
\end{center}

Figure \ref{ptdist} shows the $P_{T\mu}$ distributions for 531 KINKs 
and 276 VEEs which survived the above selection criteria.
An excess of 19 events can be seen in the region ($P_{T\mu} > 0.865 GeV/c$)
exceeding the kinematic limit of the 
$D^+ \rightarrow \bar{K}^0 \mu^+ \nu_\mu$ decay for the KINK events, 
while no such signal is seen for the VEE events.
These large $P_{T\mu}$ events can be interpreted as the fully leptonic 
decay $D_s^+ \rightarrow \mu^+ \nu_\mu$.

%

In addition to $D_s^+ \rightarrow \mu^+ \nu_\mu$ this sample includes 
background events. The largest source of background is the muon from 
the decay $D^+ \rightarrow \mu^+ \nu_\mu$. The inclusive production 
cross section for $D^+$ is a factor of two larger than $D_s$. 
However, the branching fraction for this decay is considerably smaller 
than $D_s^+ \rightarrow \mu^+ \nu_\mu$, because it is suppressed by the
ratio of CKM matrix-element, $\mid V_{cd} \mid^2 / \mid V_{cs} \mid^2$.
A muon from the cascade decay, $D_s^+ \rightarrow \tau^+ \nu_\tau$ ; 
$\tau^+ \rightarrow \mu^+ \nu_\mu \bar{\nu_\tau}$ is another source of 
background. Since the decay angle of the first decay point is extremely
small ($\approx$1-2 mrad), it is often not detectable by our ordinary 
emulsion scanning method~[\cite{scannim}]. Therefore, only the second 
decay $\tau^+ \rightarrow \mu^+ \nu_\mu \bar{\nu_\tau}$ is observable 
and therefore will appear in the emulsion as a muonic KINK.  A final
possible background is a feedthrough muon from the daughter track 
of a hadronic charm decay.

We used a Monte Carlo simulation to estimate the total number of 
background events. Charged and neutral charmed particles were generated,
using the production distribution parametrization 
$d^2\sigma / dx_F dP_T^2\propto (1-\mid x_F \mid )^n\cdot exp(-bP_T^2)$
with $n=4.25\pm0.24\pm0.23$ 
and $b=0.76\pm0.03\pm0.03(GeV/c)^{-2}$~[\cite{charm}].
The relative numbers of $D^+$,$D^0$ and $D_s^+$ mesons were generated 
according to the the cross section values measured in this 
experiment~[\cite{charm},\cite{ds},\cite{pdg}].
The baryon $\Lambda_c^+$ was generated using the cross section value
obtained from other experiments~[\cite{lambda}].
The cross section numbers employed in the Monte Carlo are 
summarized in Table \ref{cross}.
Decays of charmed particles were simulated by JETSET 7.4 with known 
branching fractions~[\cite{pdg}], except that the ratio
$f_{D^+} / f_{D_s} = 0.90$~[\cite{lattice}] was used 
for $B(D^+\rightarrow\mu^+\nu_\mu)$.

%
%
\begin{center}
TABLE \ref{cross}
\end{center}

The number of $D_s^+ \rightarrow \mu^+ \nu_\mu$ was then extracted 
from a fit to the $P_{T\mu}$ distribution of the muonic KINK sample.
In this fit, the background normalizations and shapes were fixed  
by the Monte Carlo, while the normalization for the 
$D_s^+ \rightarrow \mu^+ \nu_\mu$ decay was allowed to float.
Fig. \ref{ptdist} a) shows the fit result for the muonic KINK decays.
The points with error bars are the experimental data, while the 
solid, dotted and dashed histograms represent 
the final fit result, the sum of the backgrounds and the 
$D_s^+ \rightarrow \mu^+ \nu_\mu$\ signal, respectively. 
The fit is in a very good agreement with the data, with 
$\chi^2 / D.F.$ is $1.029$. The fit yielded 
$23.2\pm6.0_{-0.9}^{+1.0}$ $D_s^+ \rightarrow \mu^+ \nu_\mu$\ decays.
The first error is the statistical and the second one is the systematic,
involving uncertainty of production distribution.
The detection efficiency for these decays is 14.6\%,
determined mainly from a Monte Carlo calculation.
It includes efficiency factors for trigger and reconstruction (0.817),
offline selection cut (0.416), fraction of sample scanned (0.570),
fiducial and location losses (0.956), final selection cuts (0.855),
and scanning efficiency in emulsion (0.925). The overall efficiency is
large and well-understood because of the simple kink topology,
the weak event-selection cuts, and the high scanning efficiency.
The number of the background events from 
$D^+ \rightarrow \mu^+\nu_\mu$ and $D_s^+ \rightarrow \tau^+\nu_\tau$; 
$\tau^+ \rightarrow \mu^+\nu_\mu \bar{\nu_\tau}$ in the high $P_{T\mu}$
region were estimated to be 4.5 and 0.28 respectively.
The final background, feedthrough muons from hadronic decays,
was determined to be negligible.
As a systematic check, the Monte Carlo prediction for the VEE decays
was compared with the data as shown in Fig. \ref{ptdist} b).
They are in a good agreement, which gives further confidence in our
analysis.

%
%
\begin{center}
FIGURE \ref{ptdist}
\end{center}

%

In a previous publication~[\cite{ds}], we reported on 
$18.7\pm4.9_{-0.7}^{+0.4}$ $D_s^+ \rightarrow \phi \mu^+ \nu_\mu$ 
events measured in E653 without emulsion information with an efficiency
of 1.86\%. Combining these results, we can calculate the ratio 
$B(D_s^+\rightarrow\mu^+\nu_\mu)/B(D_s^+\rightarrow\phi\mu^+\nu_\mu)$
using the following equation:
\begin{equation}
\frac{B(D_s^+\rightarrow\mu^+\nu_\mu)}
{B(D_s^+\rightarrow\phi \mu^+\nu_\mu)}=
\frac{\epsilon_{\phi \mu \nu}}{\epsilon_{\mu \nu}}\cdot
\frac{N_{\mu \nu}}{N_{\phi \mu \nu}}
=0.16 \pm 0.06 \pm 0.03, \label{branch_eq}
\end{equation}
where $\epsilon$ and $N$ are the detection efficiency and the extracted 
yield for each decay channel respectively.  
The first error is the statistical and the second one is the systematic.
Using the measured value of
$B(D_s^+ \rightarrow \phi l^+ \nu_l)=1.88\pm0.29\%$~[\cite{pdg}],
we obtain,
\begin{equation}
B(D_s^+\rightarrow\mu^+\nu_\mu)=(0.30 \pm 0.12 \pm 0.06 \pm 0.05)\% 
\label{branch}
\end{equation}
 The first two errors are the statistical and systematic errors
of our analysis, involving uncertainty of production cross section 
and its distribution, and the third one reflects the uncertainty in
the branching fraction for $D_s^+ \rightarrow \phi l^+ \nu_l$.
Note that this normalization does not require knowledge of the
$D_s^+$ cross section. Furthermore, it is quite insensitive to the
$D_s^+$ production parameters $n$ and $b$ because the muon $P_T$
relative to the beam direction in $D_s^+ \rightarrow \mu^+ \nu_\mu$
is dominated by the decay $P_T$ of the muon rather than by production
$P_T$ of the $D_s^+$.

The pseudoscalar decay constant $f_{D_s}$ can be extracted from 
the following equation,
\begin{equation}
B(D_s^+\rightarrow\mu^+\nu_\mu)=
\frac{G_F^2}{8\pi}f_{D_s}^2\tau_{D_s}m_{D_s}m_\mu^2
\mid V_{cs}\mid^2(1-\frac{m_\mu^2}{m_{D_s}^2})^2, \label{fds_eq}
\end{equation}
where $G_F$ is the Fermi constant, $\tau_{D_s}$ and $m_{D_s}$
are the mean lifetime and mass of the $D_s^+$, $m_\mu$ is the mass 
of muon, and $V_{cs}$ is the Kobayashi-Maskawa matrix element.
Using $\tau_{D_s}=4.67\pm0.17\times 10^{-13}s$,
$m_{D_s}=1968.5\pm0.6\ MeV/c^2$,
$\mid V_{cs}\mid=0.9745\pm0.0007$~[\cite{pdg}],
and our result for $B(D_s^+\rightarrow\mu^+\nu_\mu)$,
we obtain,
\begin{equation}
f_{D_s}=194 \pm 35 \pm 20 \pm 14 \ MeV \label{fds}
\end{equation}
The first two errors are the statistical and systematic errors of our analysis,
and the third one reflects the uncertainty in the branching fraction for
$D_s^+ \rightarrow \phi l^+ \nu_l$.

Table \ref{comparison} compares the E653 result for $f_{D_s}$ with
those from other experiments. Agreement with WA75~[\cite{wa75}]
is excellent. Our result is also consistent at the $\sim$10\%
confidence level with the lager values of $f_{D_s}$ from
CLEO~[\cite{cleo}] and BES~[\cite{bes}].
Theoretical predictions~[\cite{Suzu85},\cite{Sinh86},
\cite{Nari87},\cite{Cola91}] span the range 130 to 350 $MeV$. 
Recent calculations using QCD sum rule~[\cite{sumrule}],
independent quark model~[\cite{quark}], 
and lattice~[\cite{lattice},\cite{milc}] techniques are in
good agreement with our result.

%
%
\begin{center}
TABLE \ref{comparison}
\end{center}

In conclusion, E653 has measured the ratio of the muonic branching 
ratios for the $D_s$; 
${B(D_s^+\rightarrow\mu^+\nu_\mu)}/
{B(D_s^+\rightarrow\phi \mu^+\nu_\mu)}=0.16 \pm 0.06 \pm 0.03$. 
With the latest value for $B(D_s^+\rightarrow\phi \mu^+\nu_\mu)$,
we obtain $B(D_s^+\rightarrow\mu^+\nu_\mu)
=(0.30 \pm 0.12 \pm 0.06 \pm 0.05)\%$.
This yields a value of $f_{D_s}=194 \pm 35 \pm 20 \pm 14 \ MeV$.

%

We gratefully acknowledge the efforts of the Fermi National 
Accelerator Laboratory staff in staging this experiment.
This work was supported in part by US Department of Energy;
the US National Science Foundation;
the Japan Society for the Promotion of Science;
the Japan-US Cooperative Research Program for High Energy Physics;
the Ministry of Education,Science and Culture of Japan;
the Korea Science and Engineering Foundation;
and the Basic Science Research Institute Program,
Ministry of Education,Republic of Korea.

%

%

\begin{figure}
\caption{
Decay $P_{T\mu}$ distributions of a) 531 KINK (muonic one prong) 
events, and b) 276 VEE (muonic two prong) events.
The points with error bars are the experimental data,
and the histograms are the Monte Carlo simulations.
The vertical dotted lines show the decay $P_{T\mu}$ limits
of the main semi-leptonic modes.
In (a), the solid, dotted and dashed histograms represent the
final fit result, the sum of the backgrounds,
and the $D_s^+ \rightarrow \mu^+ \nu_\mu$ signal, respectively.
In (b), the solid line represents the inclusive spectrum of
muons from $D^0$ decay.
}\label{ptdist}
\end{figure}

%

\begin{table}
\caption{Summary of offline selection criterion
and cut requirements.}\label{selection}
\begin{tabular}{ l  l }\hline
Offline selection & $P_{T\mu Beam} > 0.8\ GeV/c$ \\
Flight length cut & $FL > 500\ \mu m$ \\
Decay $P_{T\mu}$ cut     & $P_{T\mu}$ $ > 0.28\ GeV/c$ \\
Vertex position cut(only for VEE) & Decay vertex in emulsion bulk\\
\hline
\end{tabular}
\end{table}

\begin{table}
\caption{Cross sections used for Monte Carlo background
calculations.}\label{cross}
\begin{tabular}{ l  l }\hline
$\sigma(D^\pm;x_F>0)$ & 
$8.66\pm0.46\pm1.96\ \mu b/nucleon$~[\cite{charm}]\\
$\sigma(D^0;x_F>0)$ & 
$22.05\pm1.37\pm4.82\ \mu b/nucleon$~[\cite{charm}]\\
$\sigma(D_s^\pm;x_F>0)$ & 
$5.1\pm1.3\pm1.3\ \mu b/nucleon$~[\cite{ds},\cite{pdg}]\\
$\sigma(\Lambda_c^\pm;x_F>0)$ & 
$15.3\pm1.7\pm2.6\ \mu b/nucleon$~[\cite{lambda}]\\ \hline
\end{tabular}
\end{table}

\begin{table}
\caption{Summary of published measurement
of pseudoscalar decay constant $f_{D_s}$.}\label{comparison}
\begin{tabular}{ l  l }\hline
Experiment & $f_{D_s}$ $(MeV)$\\ \hline
E653 (This work) & $194 \pm 35 \pm 20 \pm 14 $ \\
WA75~[\cite{wa75}] (1993) & $238 \pm 47 \pm 21 \pm 43$ \\
CLEO~[\cite{cleo}] (1994) & $344 \pm 37 \pm 52 \pm 42$ \\
BES~[\cite{bes}] (1995) & $430_{-130}^{+150} \pm 40$ \\ \hline
\end{tabular}
\end{table}


\begin{references}
\bibitem{wa75} S. Aoki {\it et al.}, Progress of Theoretical Physics
Vol. {\bf{89}}, No. 1, pp. (1993) 131 \\
The WA75 value was based on the 1992 PDG value of $B(D_s^+ \rightarrow
K^+ K^- \pi^+)=(3.9\pm0.4)\%$ and $B(D^0 \rightarrow \mu \nu_\mu X)=
(8.8 \pm 2.5)\%$. 
Using the new PDG~[\cite{pdg}] value of
$B(D_s^+ \rightarrow K^+ K^- \pi^+)=(4.8\pm0.7)\%$ and
$B(D^0 \rightarrow \mu \nu X)=(7.6\pm1.7)\%$,the corrected value
is $f_{D_s}=(238 \pm 47 \pm 21 \pm 43)MeV/c^2$.

\bibitem{cleo} D. Acosta {\it et al.}, 
Phys. Rev. {\bf{D 49}} (1994) 5690.

\bibitem{bes} J.Z. Bai {\it et al.}, 
Phys. Rev. Lett. {\bf{74}} (1995) 4599.

\bibitem{argus} H. Albrecht {\it et al.},
Z. Phys. {\bf{C 54}} (1992) 1.

\bibitem{e653nim} K. Kodama {\it et al.}, 
Nucl. Instr. \& Meth. {\bf{A 289}} (1990) 146.

\bibitem{scannim} K. Kodama {\it et al.}, 
Nucl. Instr. \& Meth. {\bf{B 93}} (1994) 340.

\bibitem{lattice} C.W. Bernard, J.N. Labrenz, A. Soni, 
Phys. Rev. {\bf{D 49}} (1994) 2536.

\bibitem{charm} K. Kodama {\it et al.}, 
Phys. Lett. {\bf{B 284}} (1992) 461.

\bibitem{ds} K. Kodama {\it et al.}, 
Phys. Lett. {\bf{B 309}} (1993) 483.

\bibitem{lambda} S. Barlag {\it et al.}, 
Phys. Lett. {\bf{B 247}} (1990) 113.\\
K. Kodama {\it et al.}, Phys. Lett. {\bf{B 286}} (1992) 187.

\bibitem{Suzu85} M. Suzuki, Phys. Lett. {\bf{B 162}} (1985) 392.

\bibitem{Sinh86} S.N. Sinha, Phys. Lett. {\bf{B 178}} (1986) 110.

\bibitem{Nari87} S. Narison, Phys. Lett. {\bf{B 198}} (1987) 104.

\bibitem{Cola91} P. Colangelo, G. Nardulli, and M. Pietroni, 
Phys. Rev. {\bf{D 43}} (1991) 3002.

\bibitem{sumrule} K. Schilcher, Y.L. Wu, 
Z. Phys. {\bf{C 54}} (1992) 163.

\bibitem{quark} N. Barik, P.C. Dash, 
Phys. Rev. {\bf{D 47}} (1993) 2788.

\bibitem{milc} C. Bernard {\it et al.}, 
Nucl. Phys. B, Proc. Suppl. {\bf{42}} (1995) 388.

\bibitem{pdg} L. Montanet {\it et al.}, 
Phys. Rev. {\bf{D 50}}, 1173 (1994) and
1995 off-year partial update for the 1996 edition 
(URL:http://pdg.lbl.gov/)

\end{references}
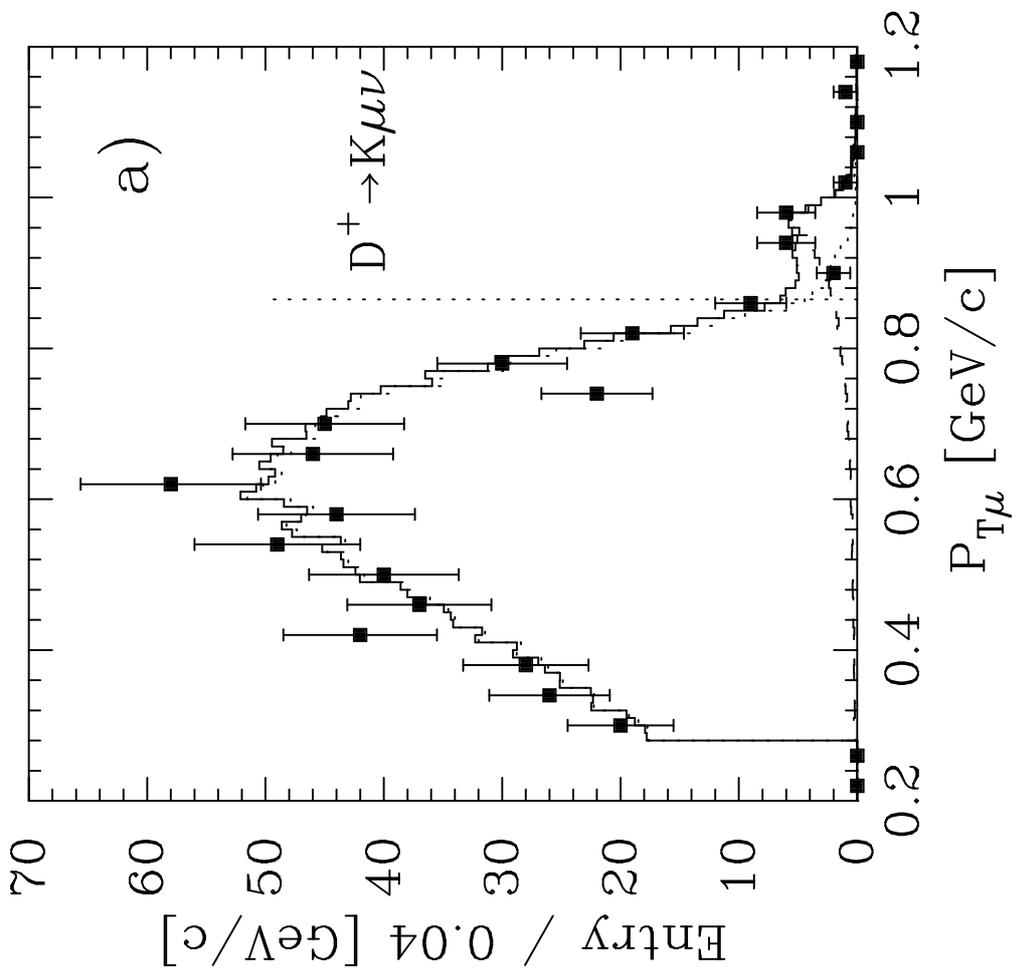
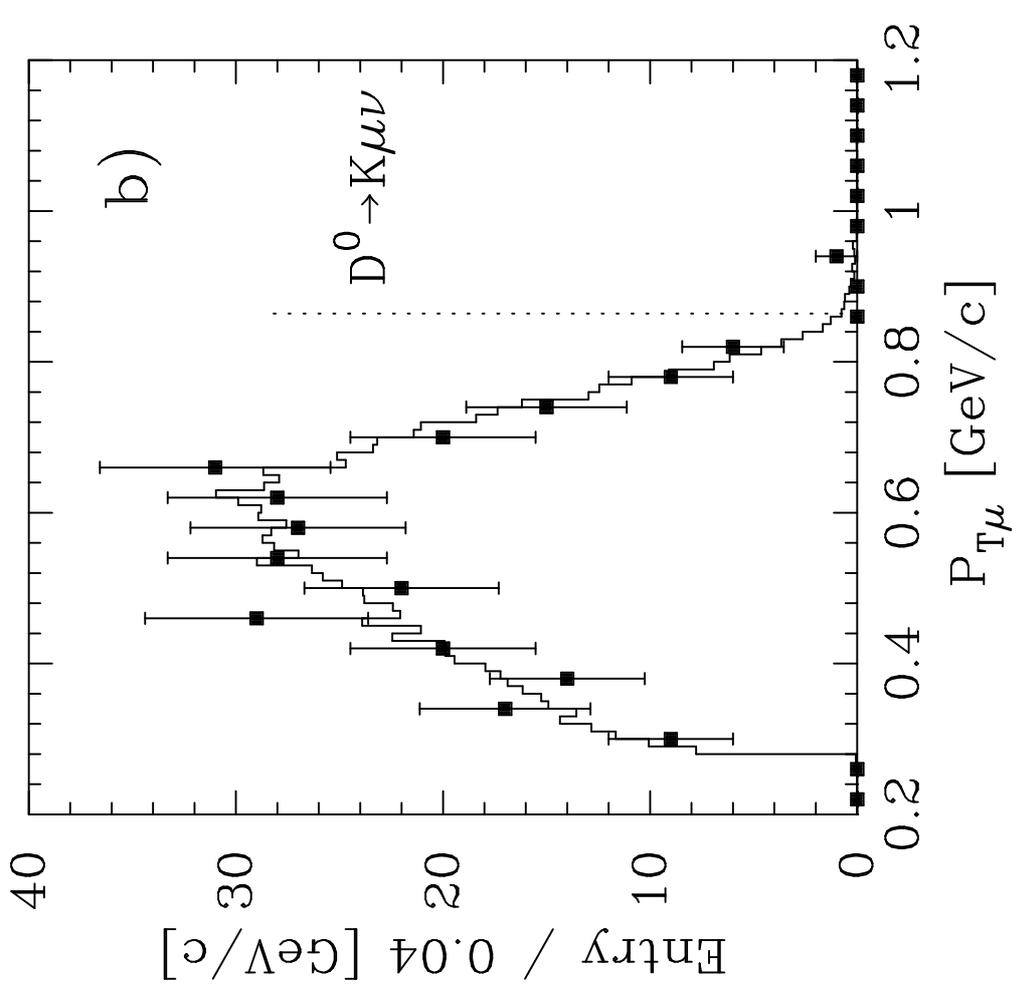
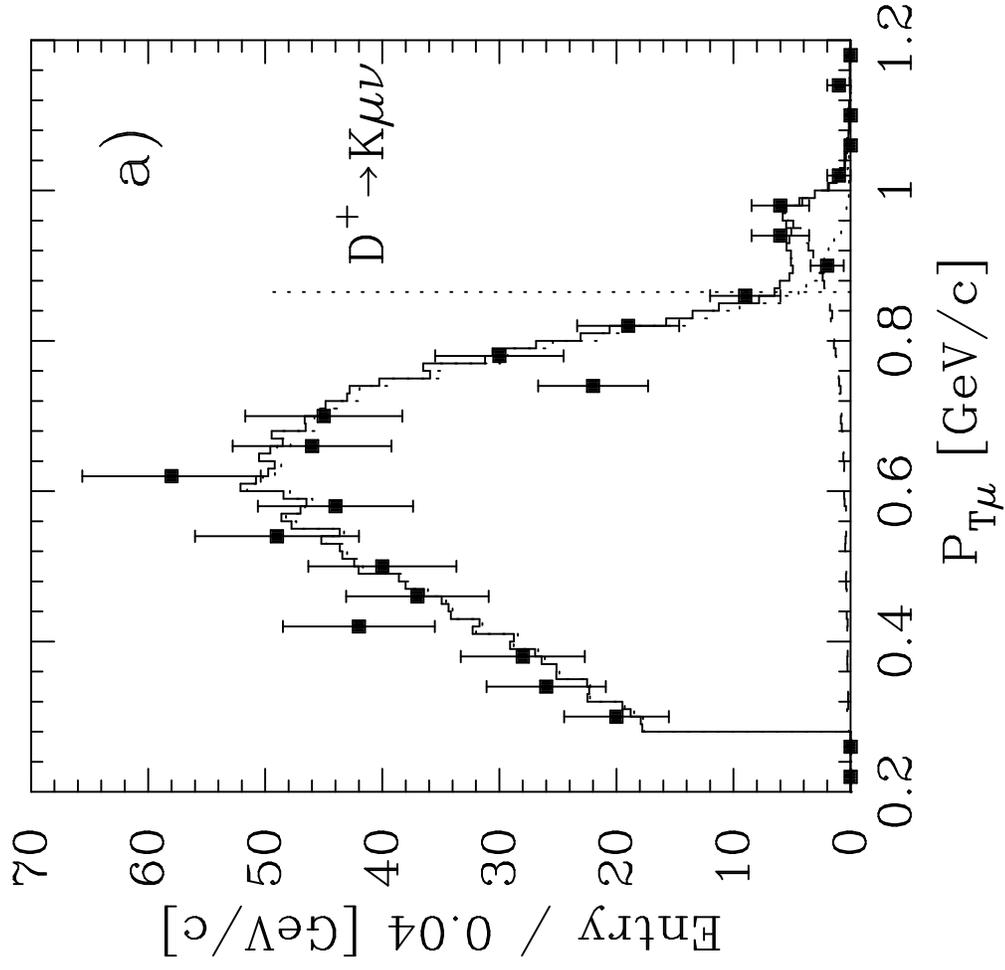
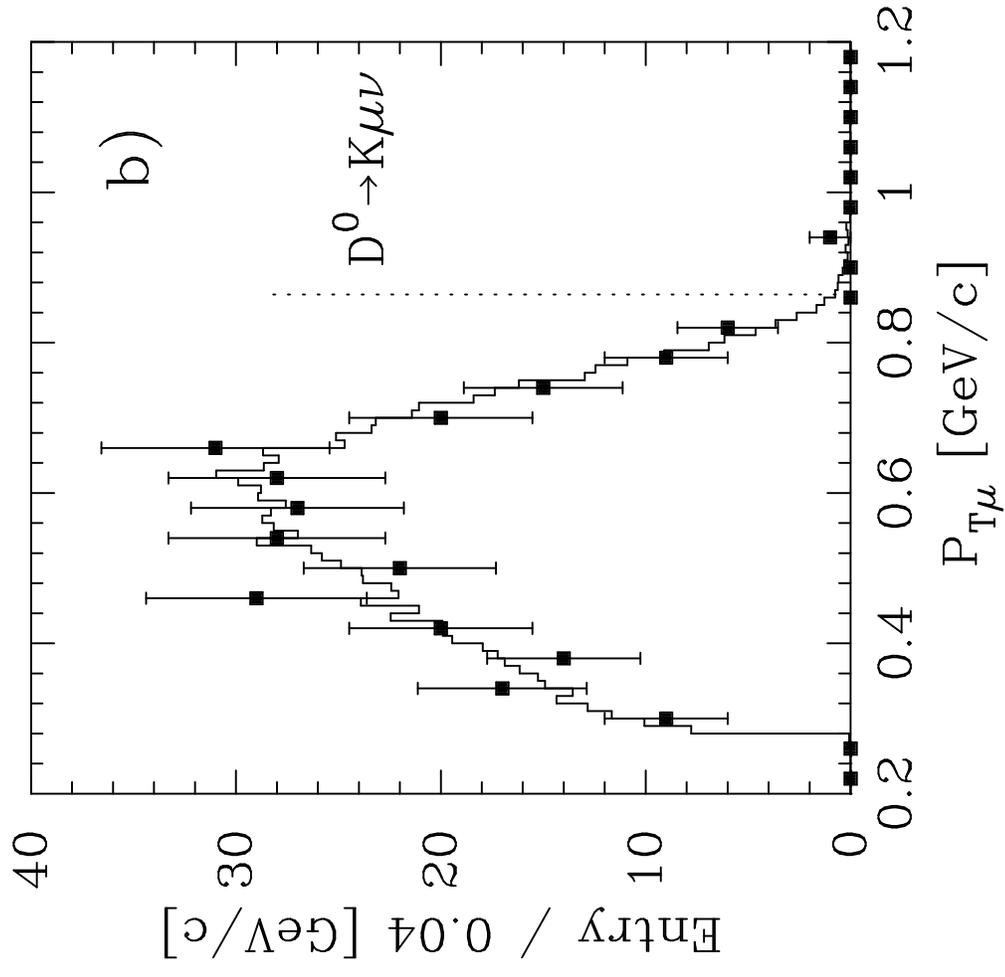
\end{document}